\documentclass[twocolumn]{aastex63}
\received{2020 December 7}
\revised{2021 January 13}
\accepted{2021 January 14}
\submitjournal{ApJL}
\shorttitle{Vortex-shedding driver}
\shortauthors{Karampelas \& Van Doorsselaere}
\graphicspath{{./}{figures/}}
\begin{document}

%\title{Decay-less loop oscillations via vortex shedding: a self oscillating process}
\title{Transverse loop oscillations via vortex shedding: a self oscillating process}

\correspondingauthor{Konstantinos Karampelas}
\email{konstantinos.karampelas@northumbria.ac.uk}

\author[0000-0001-5507-1891]{Konstantinos Karampelas}
\affiliation{Department of Mathematics, Physics, and Electrical Engineering, Northumbria University,\\ Newcastle upon Tyne, NE1 8ST, UK}
\affiliation{Centre for mathematical Plasma Astrophysics, Department of Mathematics, KU Leuven,\\ Celestijnenlaan 200B bus 2400, B-3001 Leuven, Belgium }

\author[0000-0001-9628-4113]{Tom Van Doorsselaere}
\affiliation{Centre for mathematical Plasma Astrophysics, Department of Mathematics, KU Leuven,\\ Celestijnenlaan 200B bus 2400, B-3001 Leuven, Belgium }

\begin{abstract}
Identifying the underlying mechanisms behind the excitation of transverse oscillations in coronal loops is essential for their role as diagnostic tools in coronal seismology and their potential use as wave heating mechanisms of the solar corona. In this paper, we explore the concept of these transverse oscillations being excited through a self-sustaining process, caused by Alfv\'{e}nic vortex shedding from strong background flows interacting with coronal loops. We show for the first time in 3D simulations that vortex shedding can generate transverse oscillations in coronal loops, in the direction perpendicular to the flow due to  periodic ``pushing'' by the vortices. By plotting the power spectral density we identify the excited frequencies of these oscillations. We see that these frequencies are dependent both on the speed of the flow, as well as the characteristics of the oscillating loop. This, in addition to the fact that the background flow is constant and not periodic, makes us treat this as a self-oscillating process. Finally, the amplitudes of the excited oscillations are near constant in amplitude, and are comparable with the observations of decay-less oscillations. This makes the mechanism under consideration a possible interpretation of these undamped waves in coronal loops.
\end{abstract}

\keywords{Magnetohydrodynamical simulations; Solar coronal loops; Solar coronal seismology}

\section{Introduction} \label{sec:intro}
In recent years, observations by the Coronal Multi-channel Polarimeter, the Solar Dynamics Observatory, and Hinode spacecraft have already proven the ubiquity of transverse perturbations and waves in magnetic structures in the solar corona \citep[e.g.][]{tomczyk2007, mcintosh2011}. The importance of these waves and oscillations is connected to their use in coronal seismology \citep[e.g.][]{Nakariakov2001A&A, Nakariakov2005LRSP}, as well as their potential role as heating mechanisms for the solar corona \citep[for a review, see][]{tvd2020coronal}.

Kink oscillations of coronal loops have been intensively studied ever since they were firstly observed \citep{aschwanden1999, nakariakov1999}. Following the theory of waves in a magnetized cylindrical flux tube \citep{zajtsev1975, edwin1983wave}, these observed perturbations have been treated as standing kink modes \citep{tvd2008detection}. These first observations were of oscillations with amplitudes of a few megameters, which were decaying over time after an being excited by external energetic phenomena \citep[e.g.][]{nakariakov1999,ZimovetsNakariakov2015A&A,Nechaeva2019ApJS}. The damping of these oscillations has been attributed to the phenomena of resonant absorption and phase mixing \citep{Ionson1978ApJ, heyvaerts1983,goossens2011resonant} and have been studied both analytically and numerically in 3D MHD setups, where the effects of gravity, radiation, and the Kelvin-Helmholtz instability (KHi) has also been considered \citep[e.g.][]{terradas2008, antolin2014fine, magyar2015, Hillier2019MNRAS}. 

Alongside those larger-amplitude decaying oscillations, a second category of low-amplitude, transverse waves occurring in coronal loops has also been observed in recent years. These waves were first detected by \citet{wang2012} and \citet{tian2012}, and were proven to be omnipresent in active region coronal loops \citep{anfinogentov2013, anfinogentov2015}, making them possible tools for coronal seismology \citep{anfinogentov2019ApJ,Yang2020Sci}. These decay-less oscillations have a near constant amplitude over the course of many periods, with frequencies equal to that of the fundamental standing kink mode \citep{nistico2013}, as well as its second harmonic \citep{duckenfield2018ApJ}.

While the mechanism exciting these decay-less oscillations is still not identified, different explanations have been studied over the years. In \citet{antolin2016}, they were treated as line-of-sight effects created by the KHi vortices from impulsively oscillating coronal loops. Decay-less oscillations have also been modeled numerically as driven standing waves from footpoint drivers, both from monoperiodic \citep[e.g.][]{karampelas2017,karampelas2019amp,afanasev2019,mingzhe2019,mijie2021ApJL} and broadband drivers \citep{afanasyev2020decayless}. Another interpretation was considered in \citet{nakariakov2016}, in the form of a self-sustained oscillation. Unlike periodically driven oscillations, where the input of energy is done periodically and the frequency is imposed by the driver, self-oscillations are excited from (near) constant drivers, and the oscillation frequency is set by the system itself, and not the external driver. In short, self-oscillations are processes that can turn a nonperiodic driving mechanism into a periodic signal. Such a process was modeled in \citet{karampelas2020ApJ}, where a slow constant flow around a loop's footpoint eventually led to the excitation of a weak oscillation. Despite the numerical limitations, that study has provided a first proof-of-concept for this process in a 3D simulation.

In the current work we will continue to explore the concept of self-oscillations of coronal loops, by considering the excitation of a transversely polarized wave through the mechanism of Alfv\'{e}nic vortex shedding. Evidence of vortex shedding in the solar corona was reported in \citet{samanta2019}, in the vicinity of a shrinking loop in a post-flare region. Vortex shedding due to solar wind has also been proposed by \citet{nistico2018A&A}, to explain the observed oscillations of cometary plasma tails. In a 0D model first proposed by \citet{nakariakov2009}, it was described how vortices generated by an upflow passing by a loop can excite an oscillation through a quasi-periodic horizontal force.  The phenomenon of Alfv\'{e}nic vortex shedding has already been explored numerically for a bluff body in 2D in \citet{gruszecki2010}. However, a coronal loop will behave differently than a fixed and rigid bluff body, thus making a 3D study essential. In this work we will study for the first time the excitation of kink-mode oscillations by vortex shedding for a full 3D setup of a coronal loop. We will focus on the decay-less regime, although this method can also be applied to decaying oscillations. We will show that the mechanism under consideration can be a possible interpretation of these undamped loop oscillations. Finally, we will explore how this mechanism is affected by the characteristics of both the flow and the oscillator, making it essentially a self-oscillation that can be initiated in solar coronal loops.

\section{Numerical setup} \label{sec:setup}
\begin{figure}[t]
    \centering
    \includegraphics[trim={0.cm 0.cm 0.cm 0.cm},clip,scale=0.37]{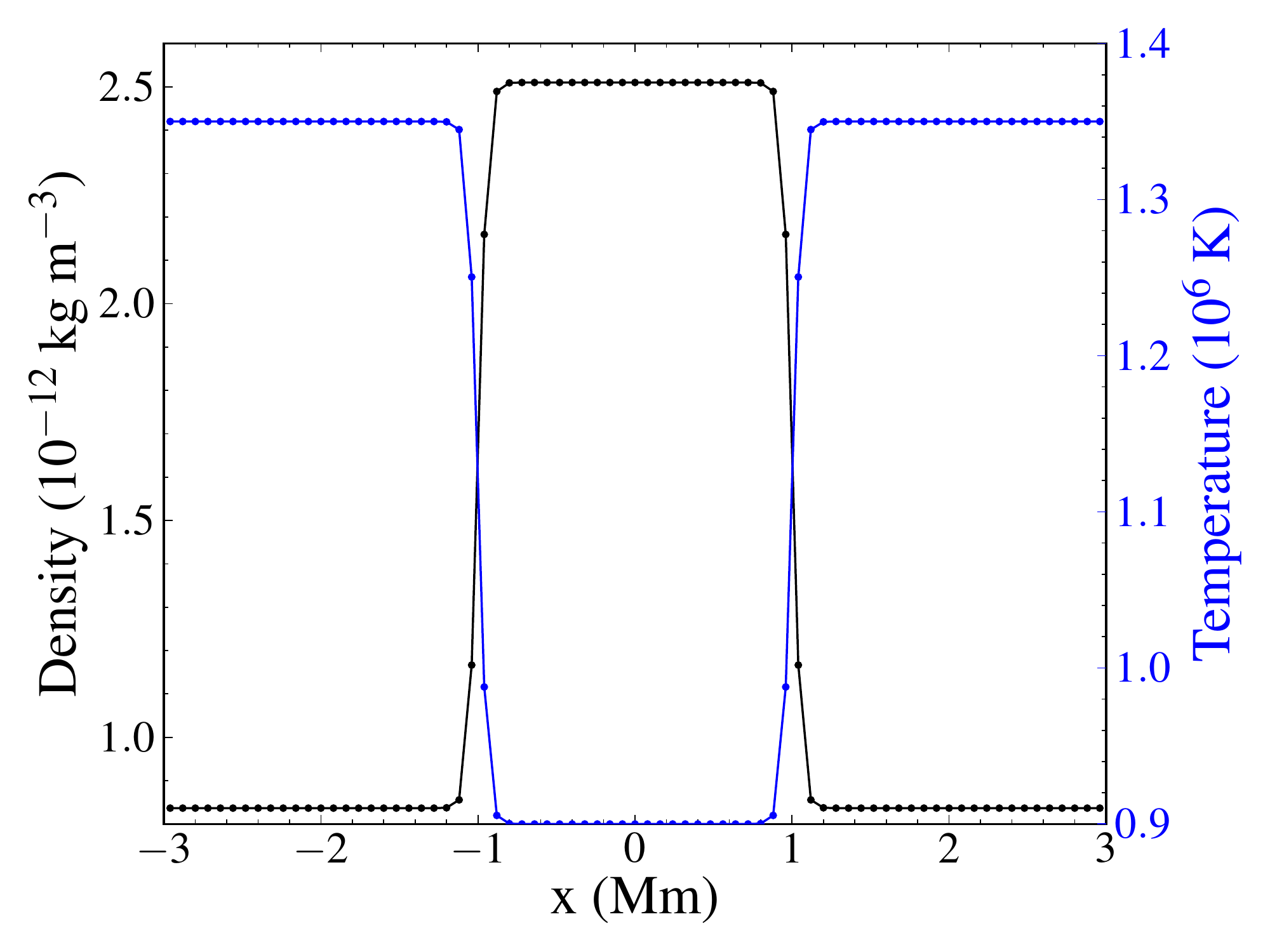}
    \caption{Density (in black) and temperature profile (in blue) of the flux tube cross-section along the $x$-axis at time $t=0$. The grid points are depicted as dots on the two curves.}
    \label{fig:profile}
\end{figure}

We use a model of a straight flux tube of (minor) radius $R=1$\,Mm and length $L=200$\,Mm. This model corresponds to a semicircular coronal loop with a major radius or $\sim 64$\,Mm. The radial density profile for our model is given by the relation
\begin{equation}
\rho(x,y) = \rho_e  + (\rho_i - \rho_e)\zeta(x,y), 
\end{equation}
\begin{equation}
\zeta(x,y) = 0.5(1-\tanh((\sqrt{x^2+y^2}/R-1)\,b)),
\end{equation}
where $b$ sets the width of the boundary layer. We consider $b=20$, which gives us an inhomogeneous layer of width $\ell \approx 0.3 R$. The index $i (e)$ corresponds to the internal (external) values with respect to our flux tube. The coronal background (or external) density is equal to $\rho_e=0.836\times 10^{-12}$\,kg\,m$^{-3}$, three times lower than the loop (internal) density ($\rho_i$). Similarly to \citet{karampelas2020ApJ}, the temperature varies across the tube axis (in the $xy$-plane), ranging from $0.9$\,MK inside the loop to $1.35$\,MK outside (see Figure \ref{fig:profile}), effectively modeling a loop during a cooling phase, as observed for loops in thermal nonequilibrium \citep{froment2017tne}. The scale height for our model ($H\sim 55$\,Mm) is comparable to the major radius or the corresponding coronal loop, which allows us to approximate the temperature as constant with height, along the flux tube. For our primary setup, we consider a straight magnetic field $B_z$, parallel to the loop axis ($z$-axis) with internal and external field values of $B_{zi} = 22.8$\,G and $B_{ze} = 22.9$\,G respectively. The magnetic field distribution is set in such a way that it maintains a total (magnetic + gas) pressure balance across our domain. This prevents any unwanted perturbations from developing at the start of the simulation.

Our setup has domain dimensions of $x\in [-7,17]$\,Mm, $y\in [-20,20]$\,Mm and $z\in [-100,100]$\,Mm, with a resolution of  $(\delta x,\delta y,\delta z) = (80,80,2000)$\,km. The loop footpoints are placed at positions $z=-100$ and $z=100$\,Mm, while $z=0$ is the location of the loop apex. This resolution allows us to study the motion of the loop and the development of larger scale flow instabilities, like vortex-shedding, in the $xy$-plane.

\begin{figure}[t]

    \centering
    \includegraphics[trim={2cm 5cm 2cm 6cm},clip,scale=0.26]{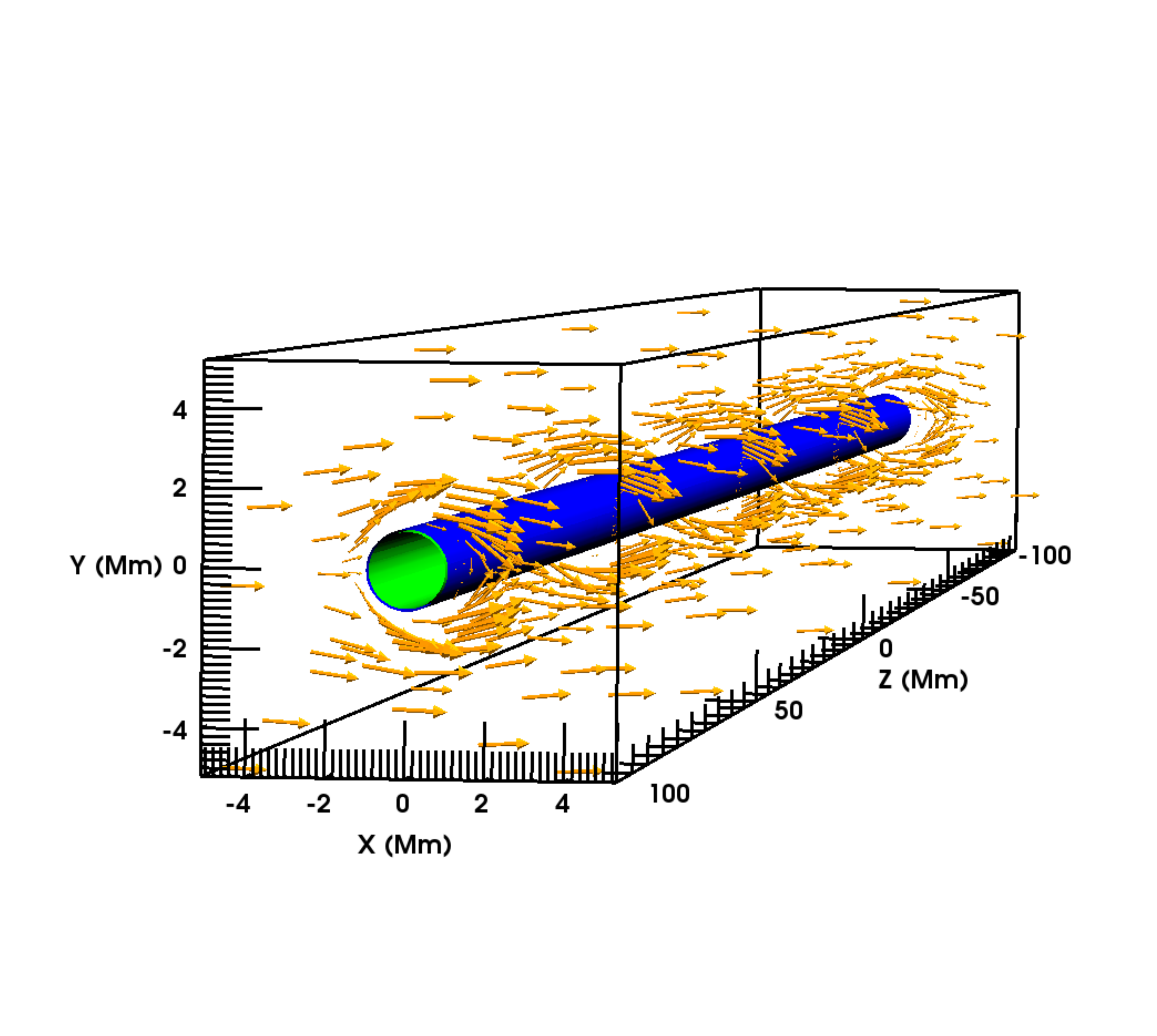}
    \caption{Schematic representation of the flow along the $x$ direction, originating from the ``left'' side boundary at $x=-7$\,Mm (not shown here).}
\label{fig:flow}
\end{figure}

The side boundaries in the $y$ direction, as well as at $x=17$\,Mm are set to have Neumann-type, zero-gradient conditions for all quantities. On the ``left'' side boundary (at $x=-7$\,Mm), we apply zero-gradient conditions for the pressure and density, the three components of the magnetic field, and the $v_y$ and $v_z$ components of the velocity field. For the $x$ velocity component at $x=-7$\,Mm, we apply a fixed value of $v_x=5\times10^4$\, m s$^{-1}$, for a total duration of $\Delta t = 1012$\,s. This leads to the development of a horizontal flow along the $x$ direction, which is also free to evolve along the $y$ direction as well (see Figure \ref{fig:flow}). After a time $t = 1012$\,s we switch the boundary condition for $v_x$ at $x=-7$\,Mm to zero-gradient, letting the initiated flow to evolve freely. 

At the ``bottom" and ``top" boundaries ($z=-100$ and $z=100$\,Mm), we apply zero-gradient conditions for the pressure, density, and the three components of the magnetic field. The $v_z$ velocity component (along the axis of the loop) is set as antisymmetric, to prevent any outflows from the top and bottom boundaries, where the bases of the loop are located. Inside the loop ($\sqrt{x^2+y^2}\leq R$), the $v_x$ and $v_y$ are set as antisymmetric, to fix the loop endpoints. Outside the loop at $z=-100$ and $z=100$\,Mm we set the $v_x$ and $v_y$ components as zero-gradient (outflow conditions) in order to let the flow evolve freely along the $xy$-plane, in a way that allows for the development of vortices.

All calculations were performed in ideal MHD in the presence of numerical dissipation, using the PLUTO code \citep{mignonePLUTO2012}. We use the second-order characteristic tracing method to calculate the timestep, and the finite volume piecewise parabolic method (PPM) with a second-order spatial global accuracy and the Roe solver. Finally, to keep the solenoidal constraint on the magnetic field, we employ Powell's $8$-wave scheme.

\section{Results} \label{sec:results}
Following the basic idea from \citet{nakariakov2009}, we try to model an upflow around a coronal loop, originating from the propagation of a CME. To that end, we initiate a flow from one of the side boundaries (at $x=-7$\,Mm) for a duration of $t=1012$\,s or $t=4\cdot P$, where $P=253$\,s is the period of the fundamental kink mode \citep{edwin1983wave} for a loop with the characteristics of our primary setup, as described in Section \ref{sec:setup}. A schematic representation of that flow, is shown in Figure \ref{fig:flow}.  Driving with a constant background flow, equal for all heights, is a rather unlikely physical scenario, which can render a straight flux tube unstable. This is due to the strong excitation of higher harmonics due to the spatial profile of the flow over the loop height. To prevent this, we ``switch off'' the side boundary driver for $t>1012$\,s and replace it with a zero-gradient boundary condition at $x=-7$\,Mm. This allows the flow to evolve freely through its interaction with the loop.

\begin{figure*}[t]
    \centering
    \resizebox{\hsize}{!}{\includegraphics[trim={0.0cm 1.9cm 4.4cm 2.cm},clip,scale=0.4]{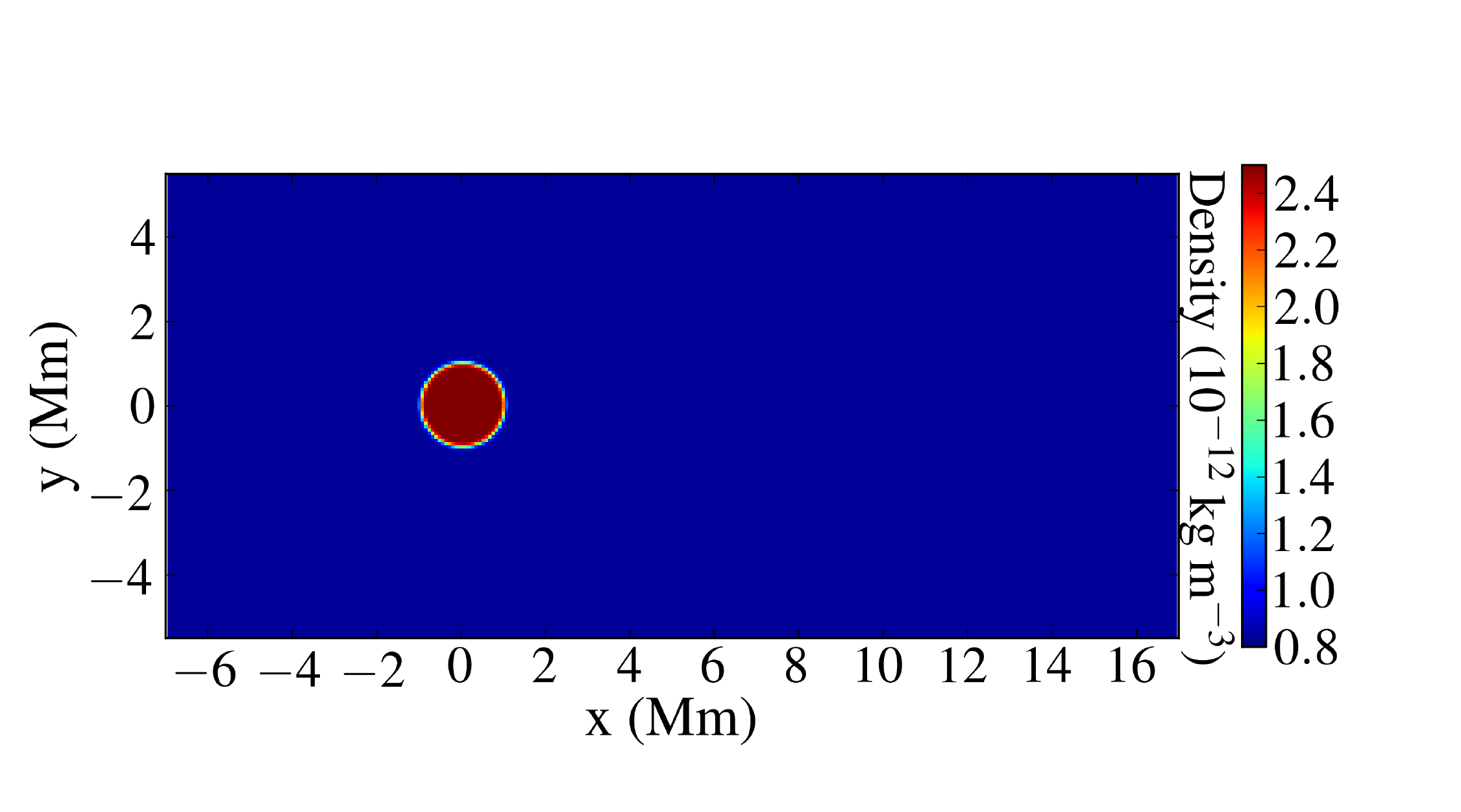}
    \includegraphics[trim={2.5cm 1.9cm 4.4cm 2.cm},clip,scale=0.4]{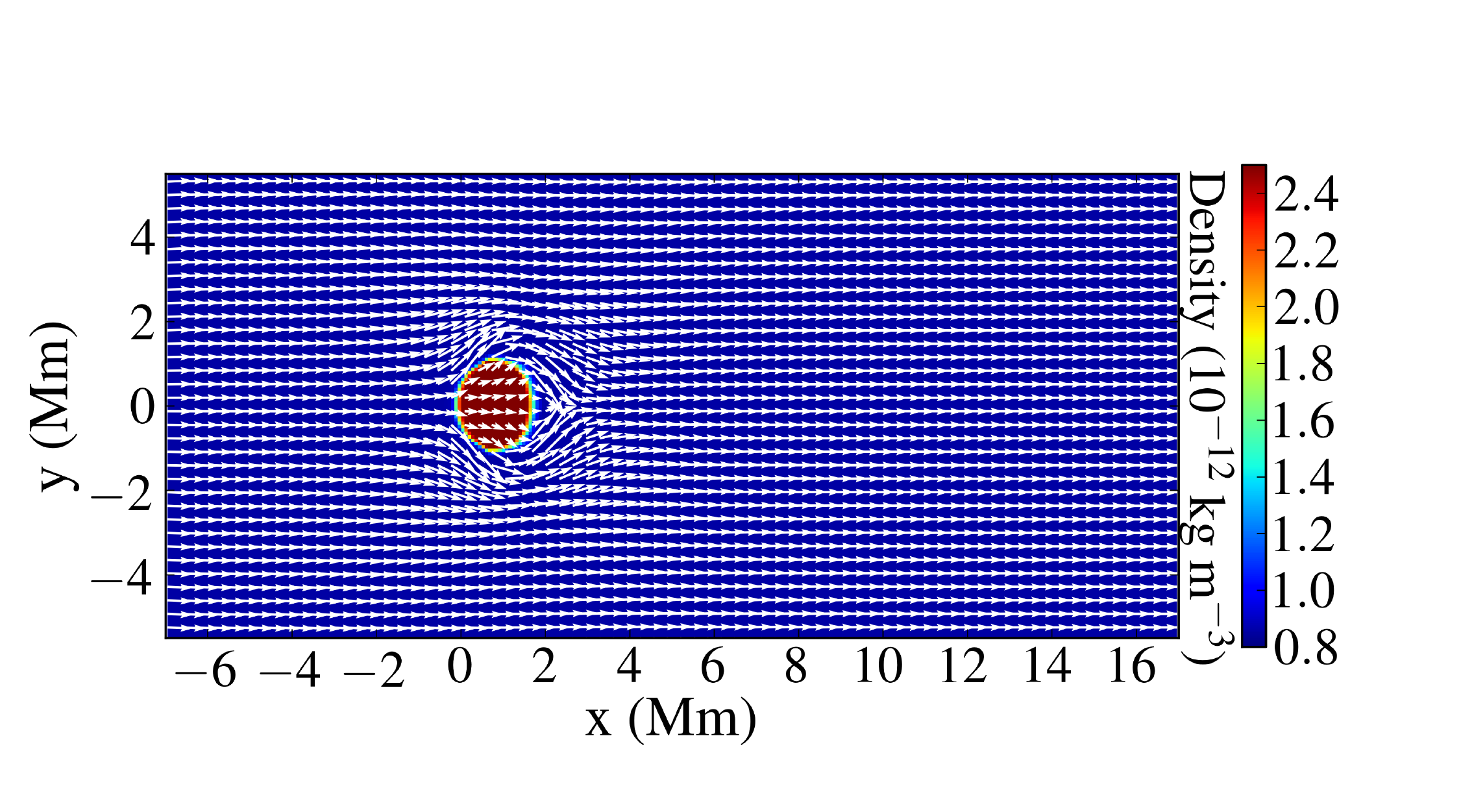}
    \includegraphics[trim={2.5cm 1.9cm 0.0cm 2.cm},clip,scale=0.4]{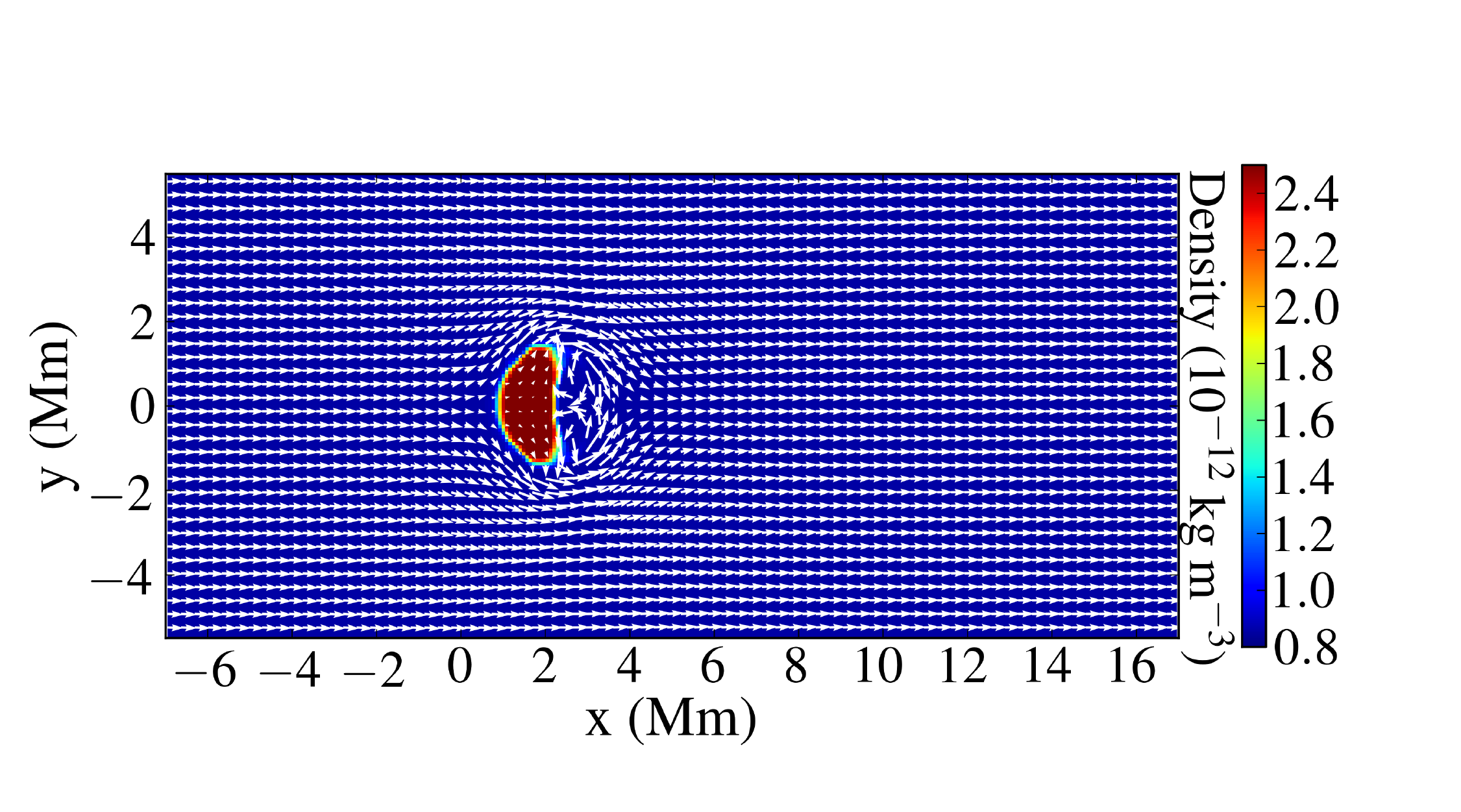}}\\
    \resizebox{\hsize}{!}{\includegraphics[trim={0.0cm 1.0cm 4.4cm 2.cm},clip,scale=0.4]{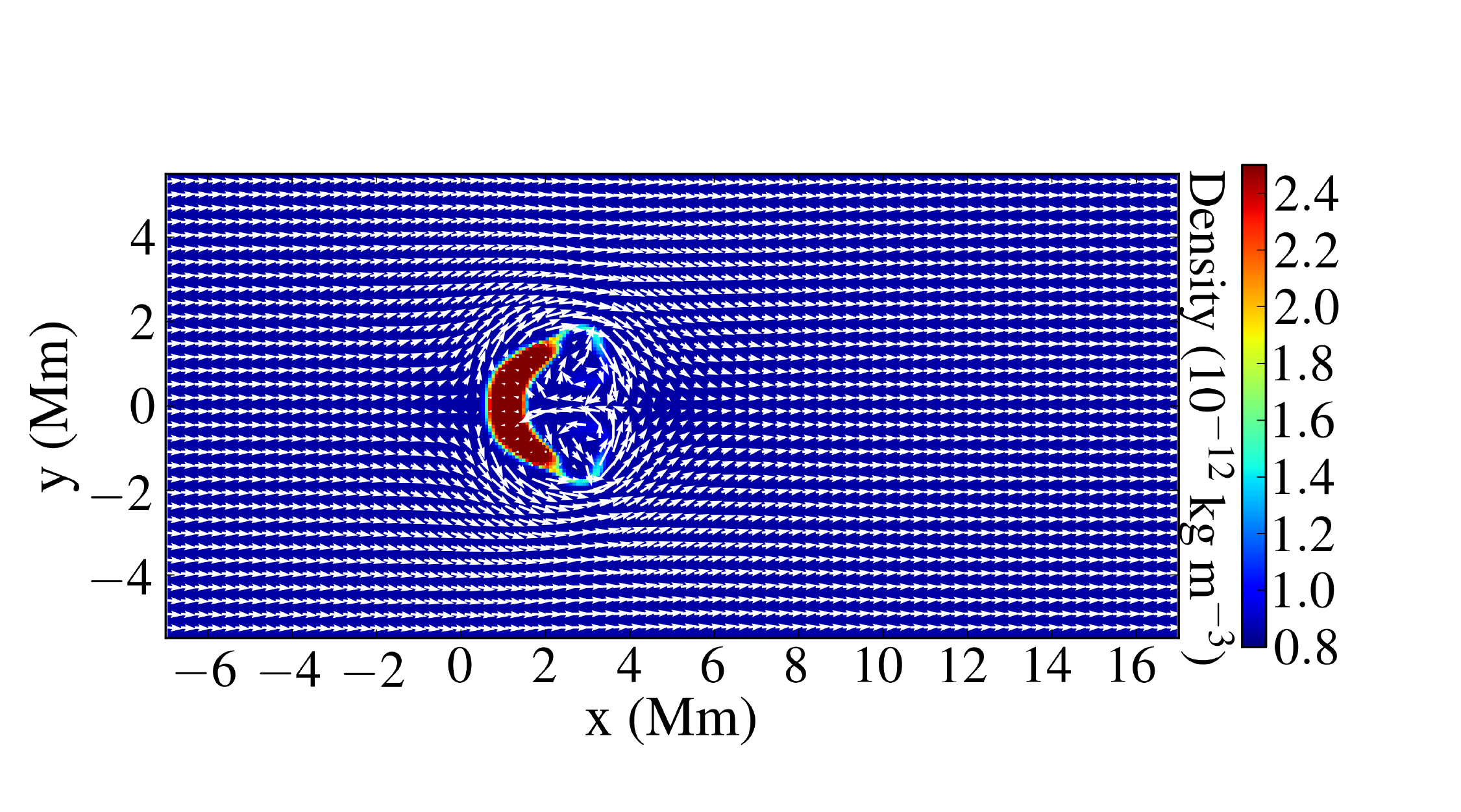}
    \includegraphics[trim={2.5cm 1.0cm 4.4cm 2.cm},clip,scale=0.4]{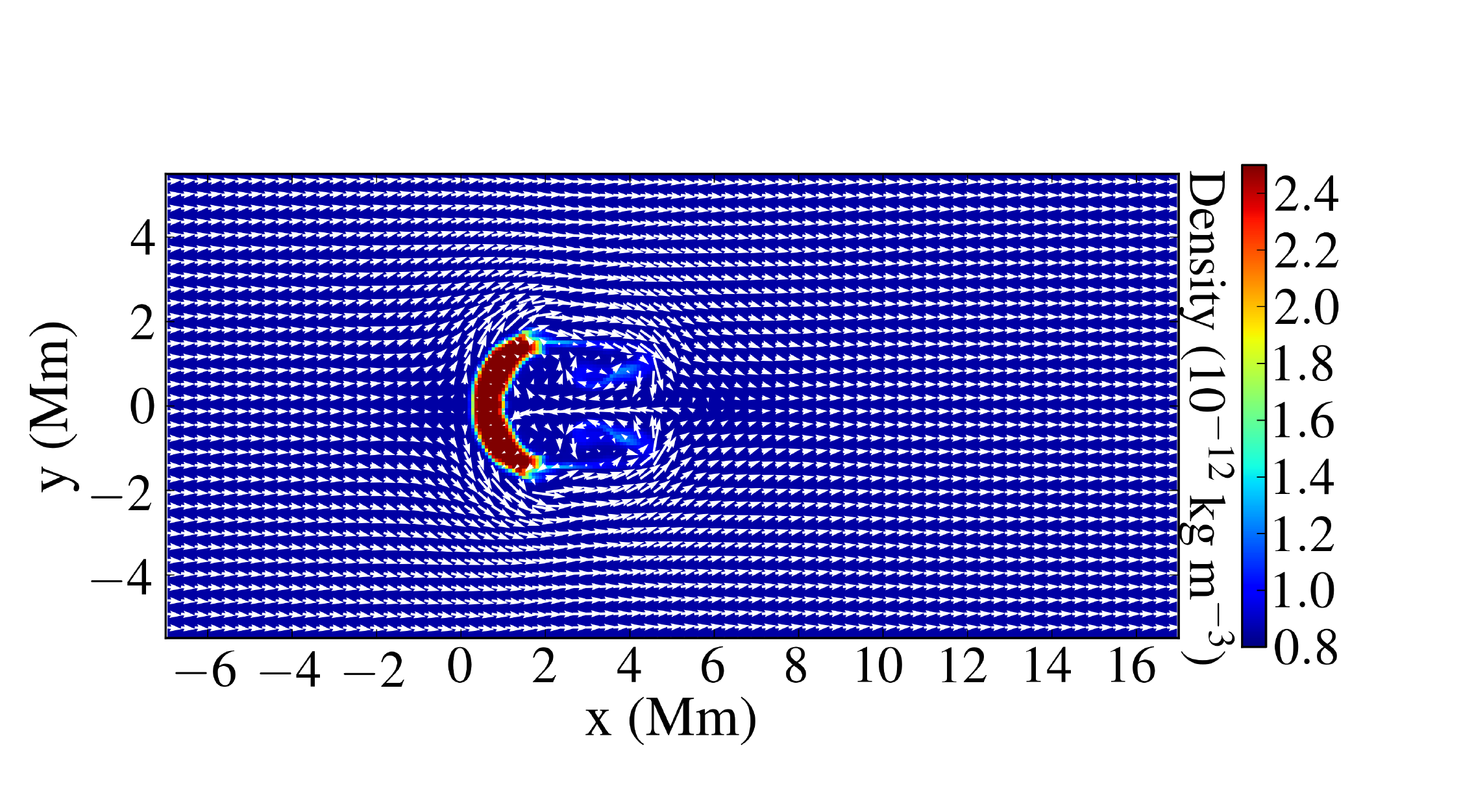}
    \includegraphics[trim={2.5cm 1.0cm 0.0cm 2.cm},clip,scale=0.4]{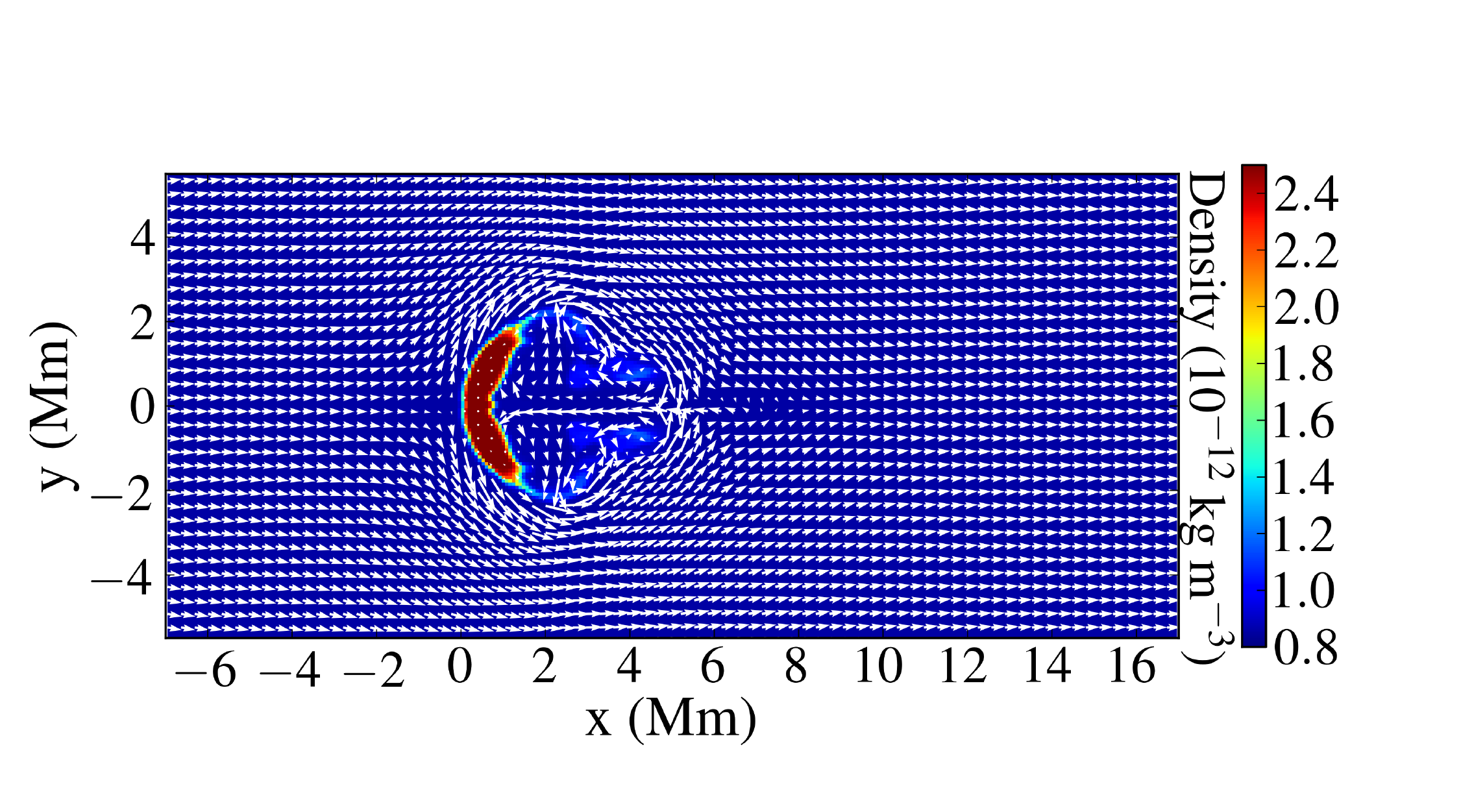}}
    \caption{Contour plots of density ($\times10^{-12}$\,kg m$^{-3}$) for our loop, at the apex. The velocity field is also overplotted. From left to right, starting from the top panels, we show the contours at the first six snapshots of the simulation, between $t=0$ and $t=189.8$\,s, for every $31.62$\,s. An animation of the density evolution is included in the online version of this manuscript.}
    \label{fig:density}
\end{figure*}

\begin{figure*}[t]
    \centering
    \resizebox{\hsize}{!}{\includegraphics[trim={0.0cm 1.9cm 4.4cm 2.cm},clip,scale=0.4]{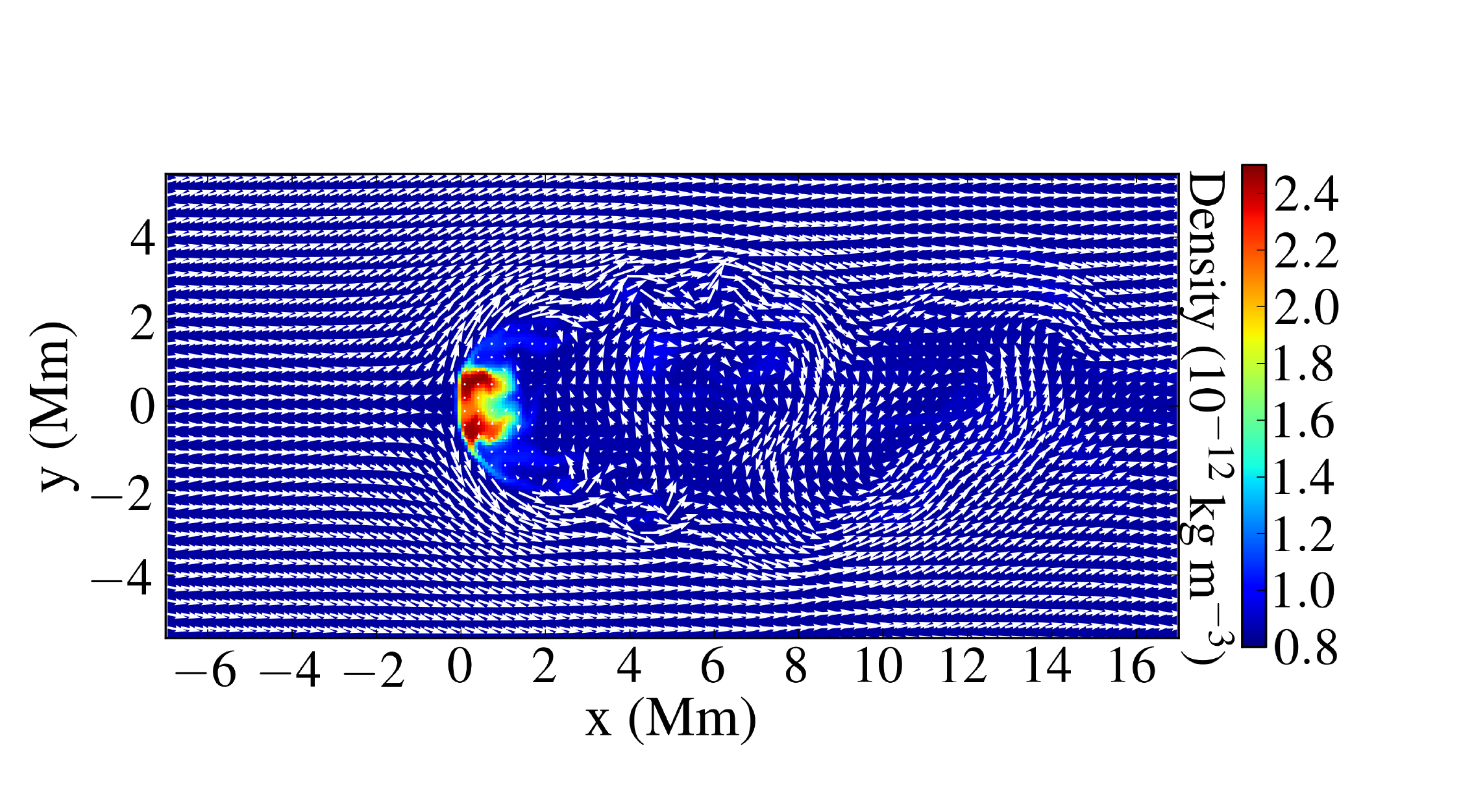}
    \includegraphics[trim={2.5cm 1.9cm 4.4cm 2.cm},clip,scale=0.4]{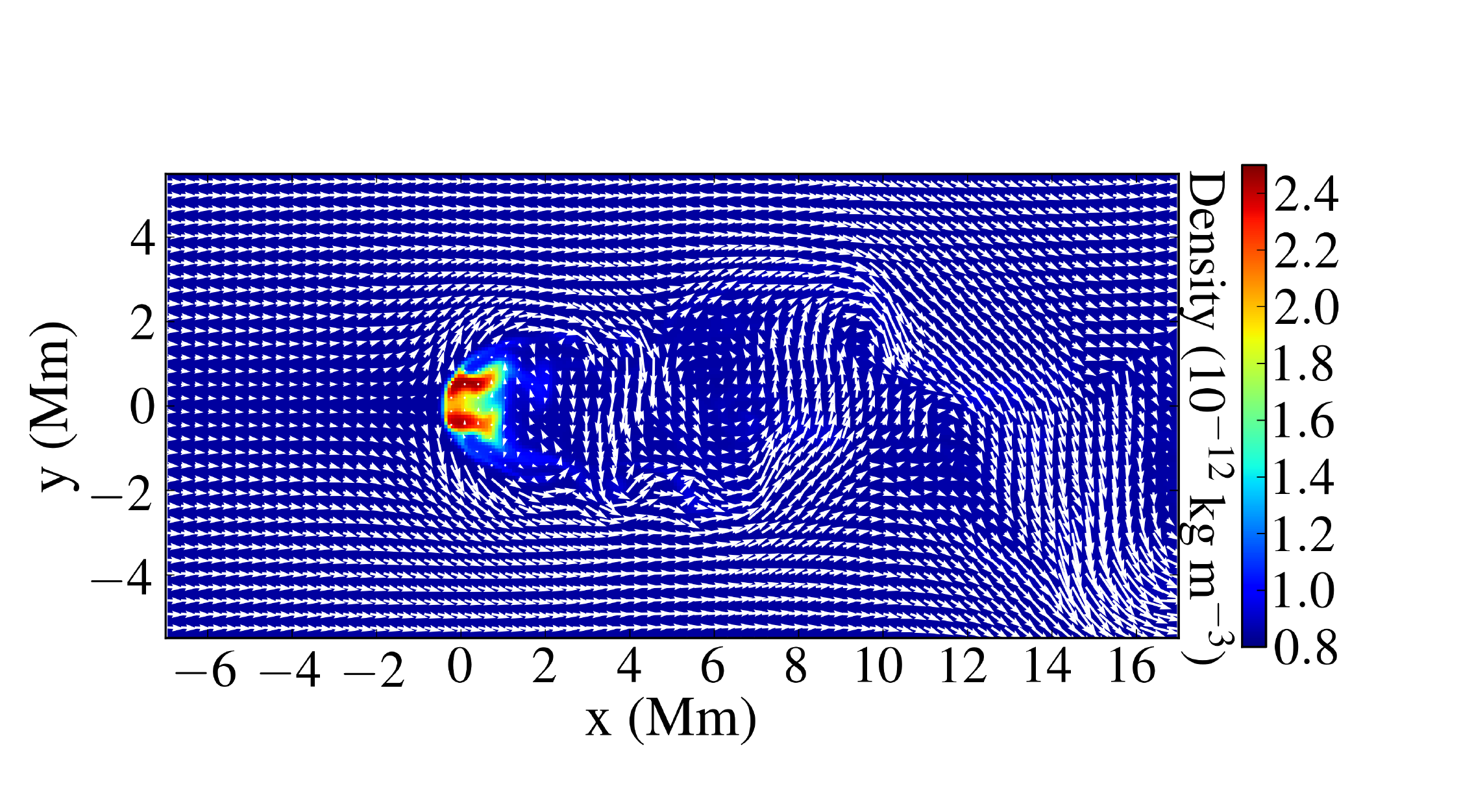}
    \includegraphics[trim={2.5cm 1.9cm 0.0cm 2.cm},clip,scale=0.4]{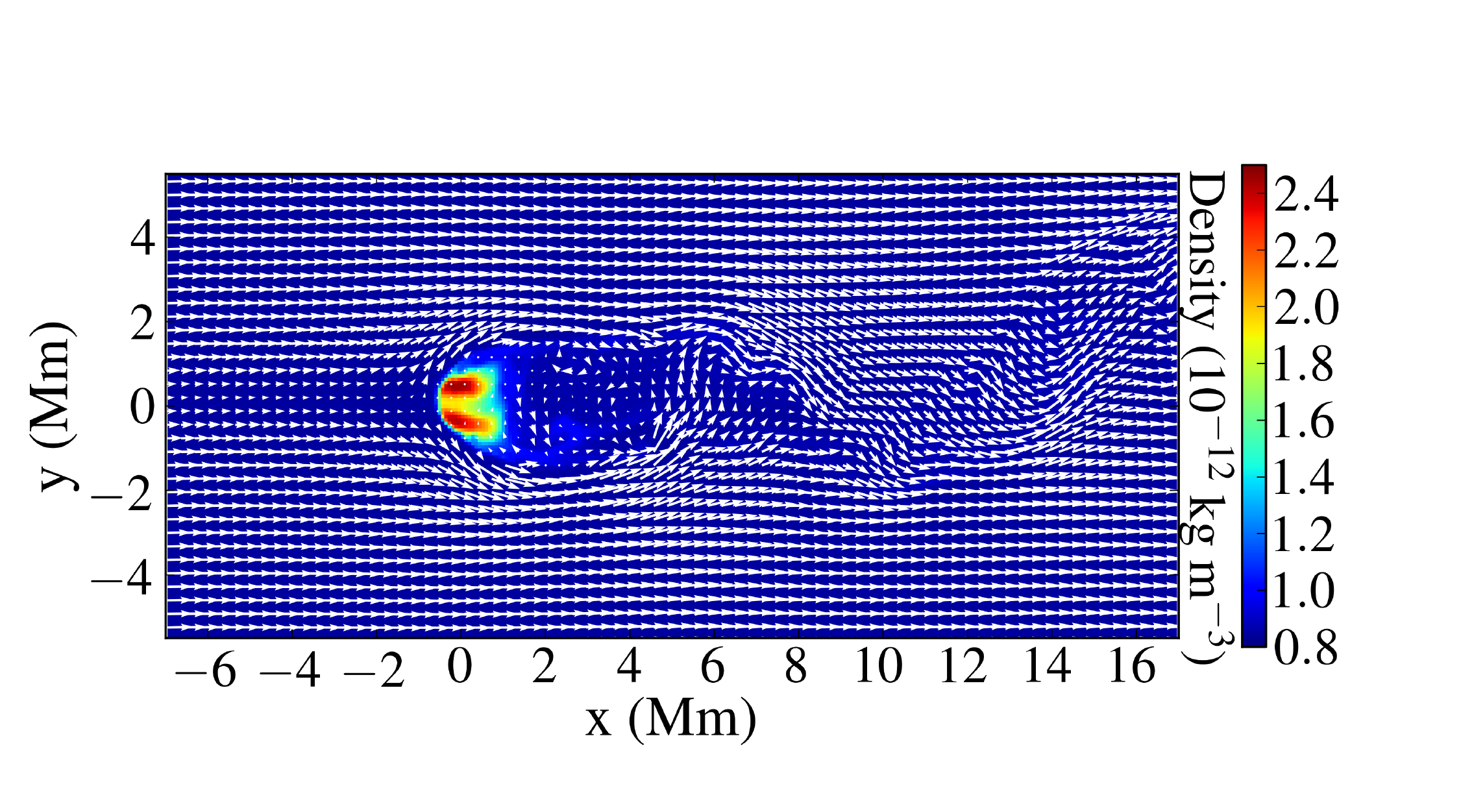}}\\
    \resizebox{\hsize}{!}{\includegraphics[trim={0.0cm 1.0cm 4.4cm 2.cm},clip,scale=0.4]{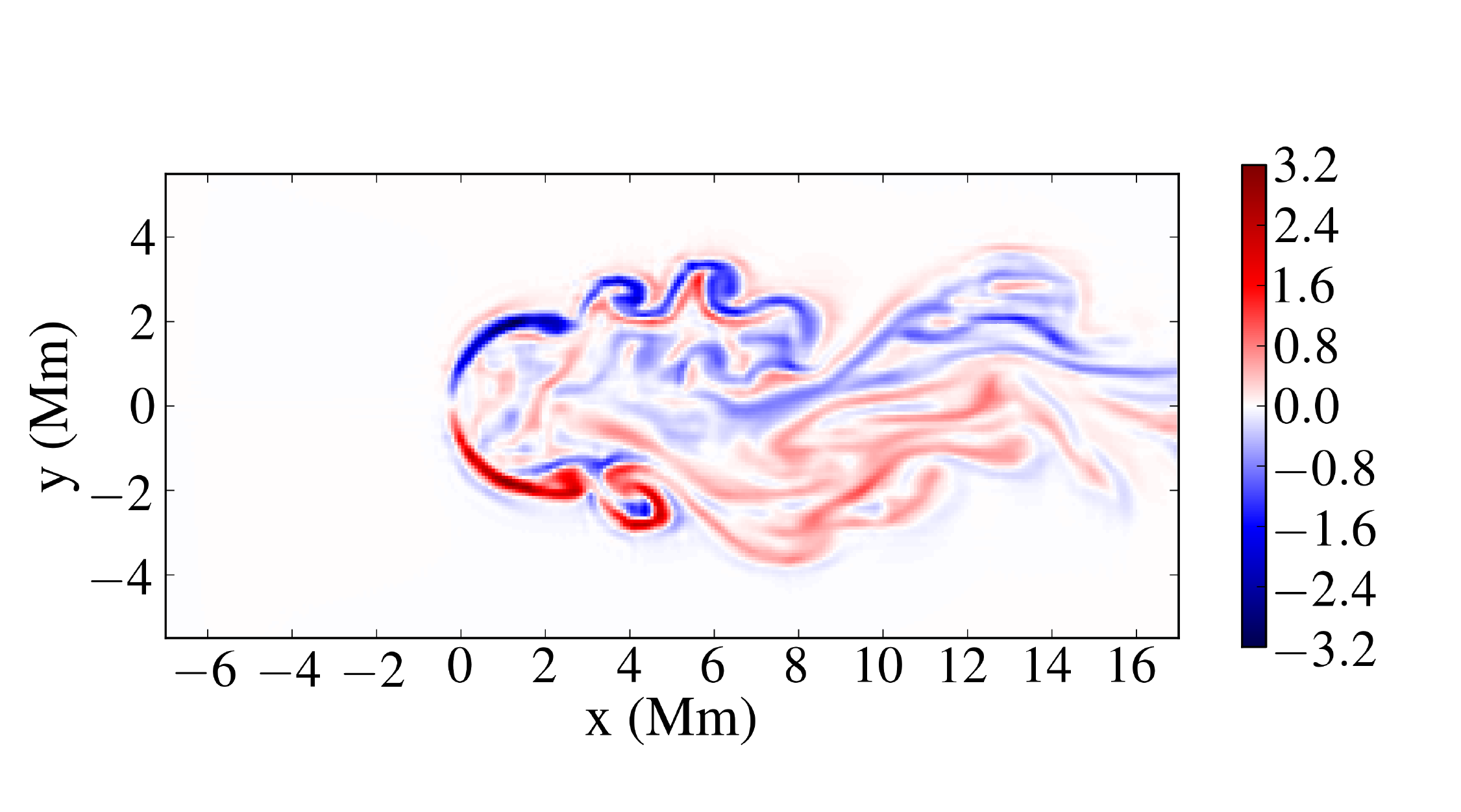}
    \includegraphics[trim={2.5cm 1.0cm 4.4cm 2.cm},clip,scale=0.4]{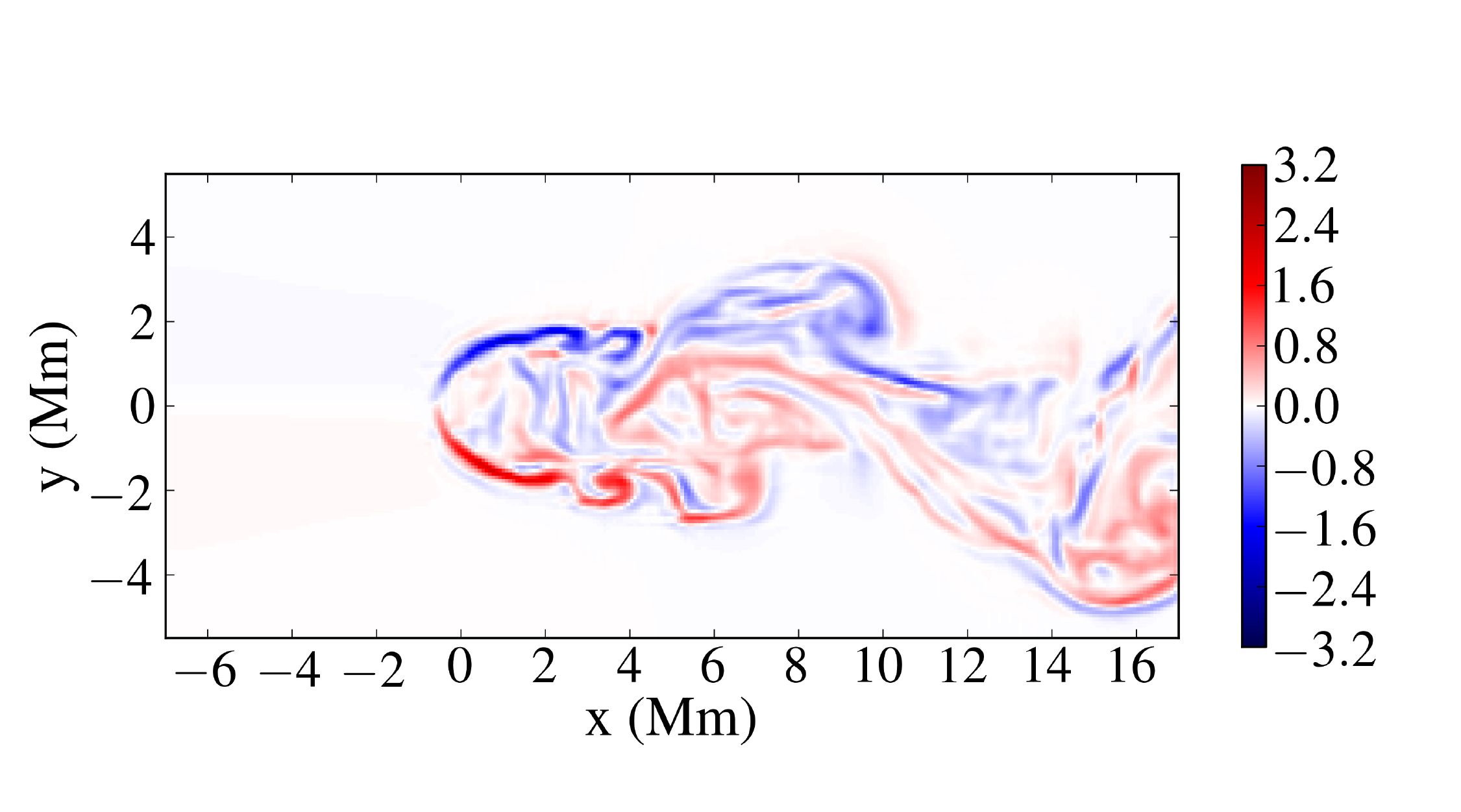}
    \includegraphics[trim={2.5cm 1.0cm 0.0cm 2.cm},clip,scale=0.4]{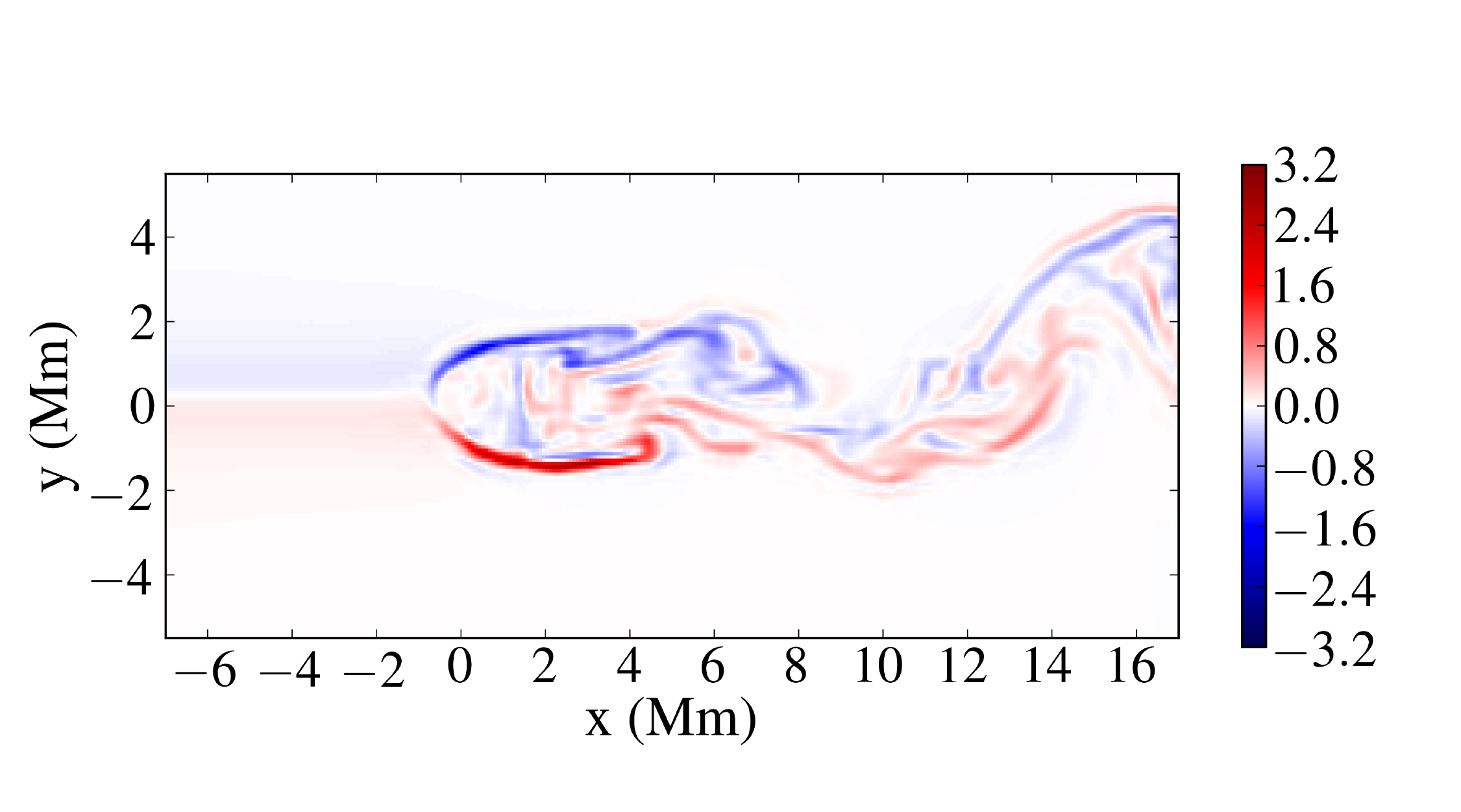}}
    \caption{Top panels: contour plots of density ($\times\,10^{-12}$\,kg m$^{-3}$) for our loop at the apex, with the overplotted velocity field in white. Bottom panels: contour plots of the plasma $z$-vorticity ($\times\, 0.1285$ Hz) for our loop at the apex. From left to right, snapshots are shown at $t=1012$, $1265$ and $1518$\,s. An animation of the $z$-vorticity ($\omega_z$) evolution is included in the online version of this manuscript.}
    \label{fig:vort}
\end{figure*}

The phenomenon of Alfv\'{e}nic vortex shedding was studied in 2D for a coronal environment in \citet{gruszecki2010}. In that study, a bluff (fixed and rigid) body was introduced in the path of a uniform plasma flow, initiating vortex shedding. The hydrodynamical relation for the Strouhal number (St) was tested, which is a dimensionless parameter depending on the period ($P$) of the vortex shedding, the size (here diameter, $d$) of the blunt body and the flow velocity ($V_0$):
\begin{equation}
    St = \frac{d}{PV_0}.
\end{equation}
In \citet{gruszecki2010} it was found that the Strouhal number in MHD for coronal parameters has values between $0.15$ and $0.25$. Considering a loop with diameter $d=2$\, Mm (minor radius of $1$\, Mm) and a period for the fundamental kink-mode equal to $253$\,s, flow speeds of $V_0 \sim 30-50$\,km s$^{-1}$ would be required for the initiation of vortex shedding with the same periodicity. This would be essential in order for the loop to resonate with and be driven by the vortices. Assuming an average value of St$\sim0.2$ for the Strouhal number, we chose to initiate a flow from the side boundaries with a velocity of $v_x=V_0=50$\,km s$^{-1}$, in order to excite an oscillation with a frequency close to that of the fundamental kink mode.

\begin{figure*}[t]
    \centering
    \includegraphics[trim={0.cm 0.cm 0.cm 0.cm},clip,scale=0.4]{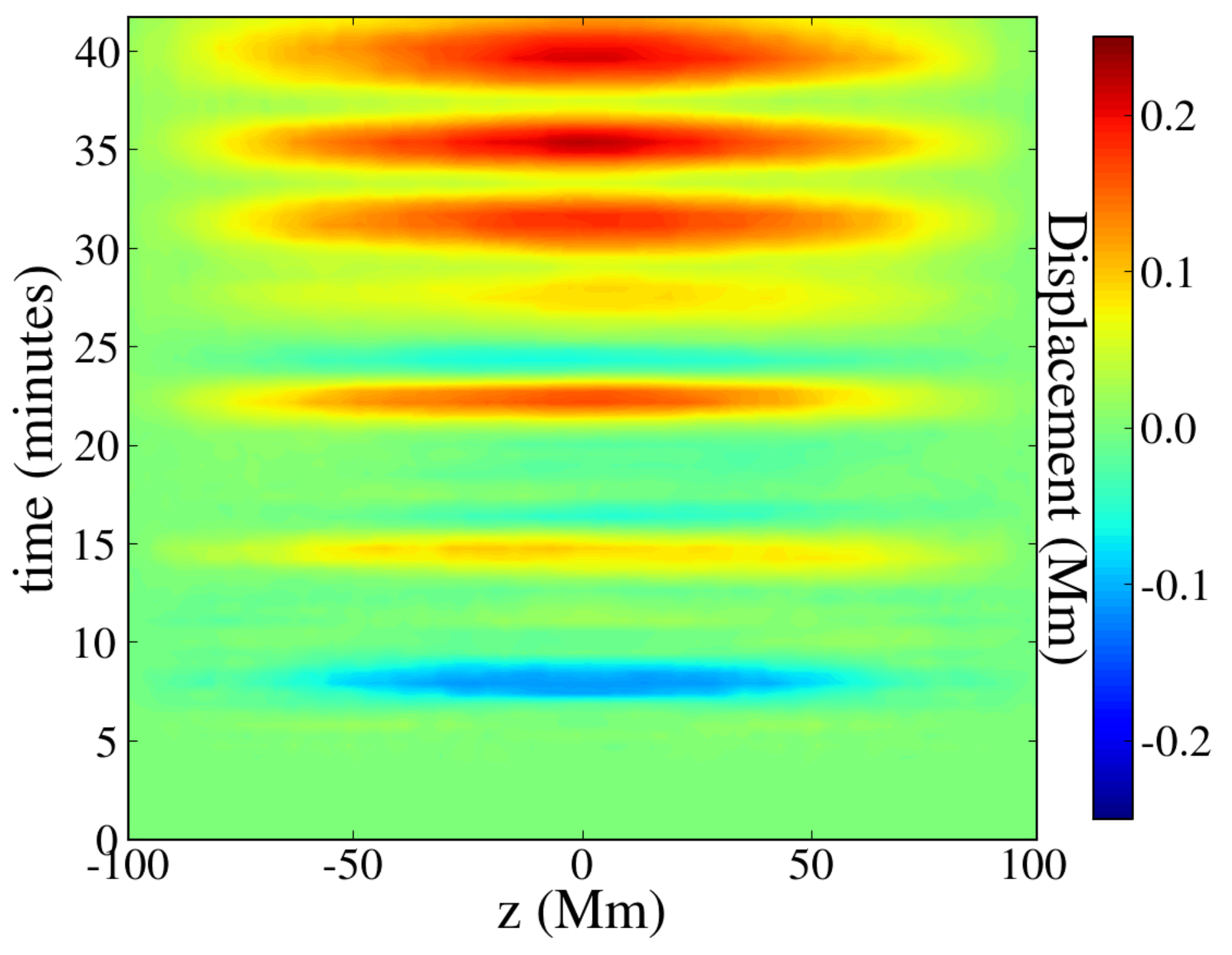}
    \includegraphics[trim={0.cm 0.cm 0.cm 0.cm},clip,scale=0.4]{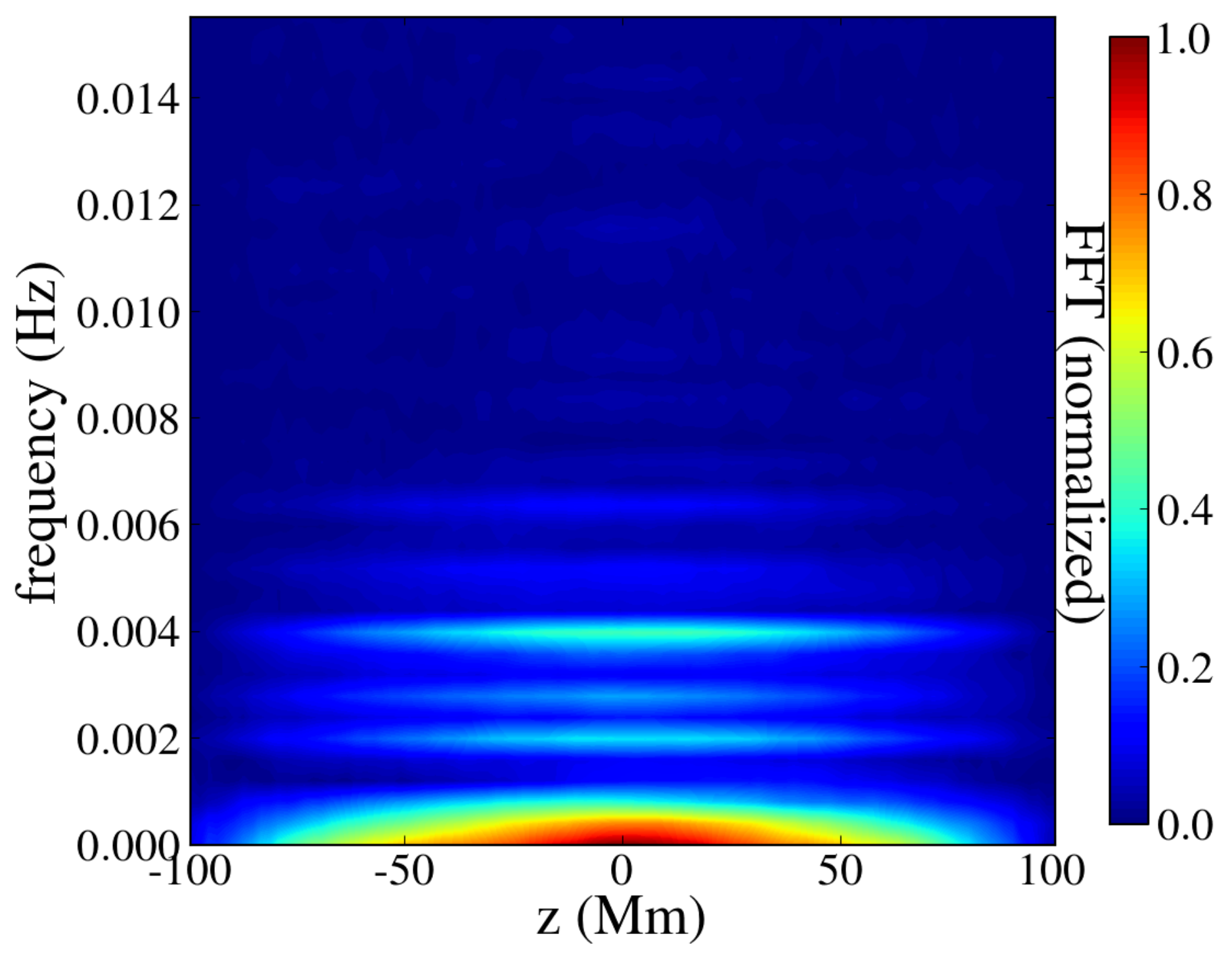}
    \includegraphics[trim={0.cm 0.cm 0.cm 0.cm},clip,scale=0.4]{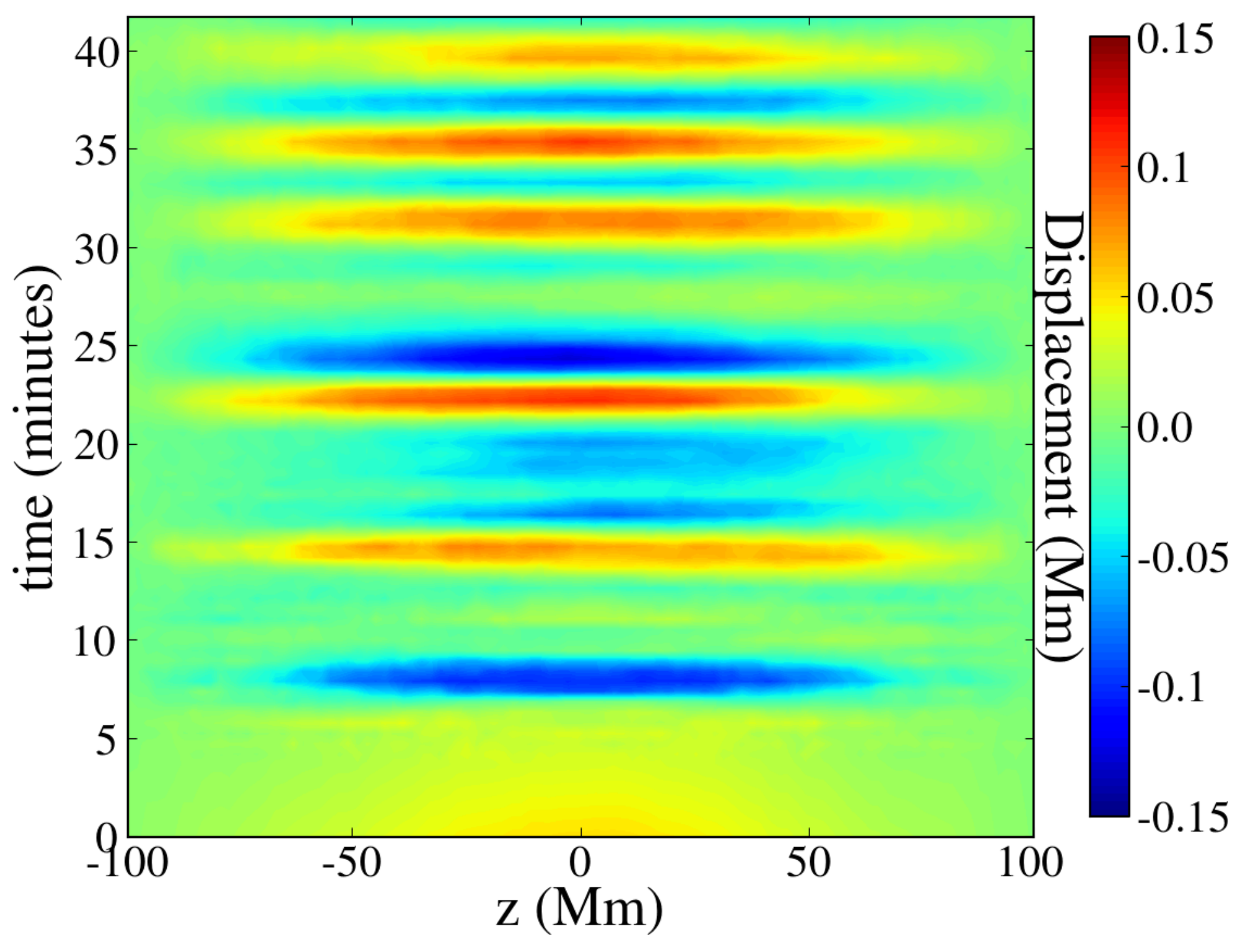}
    \includegraphics[trim={0.cm 0.cm 0.cm 0.cm},clip,scale=0.4]{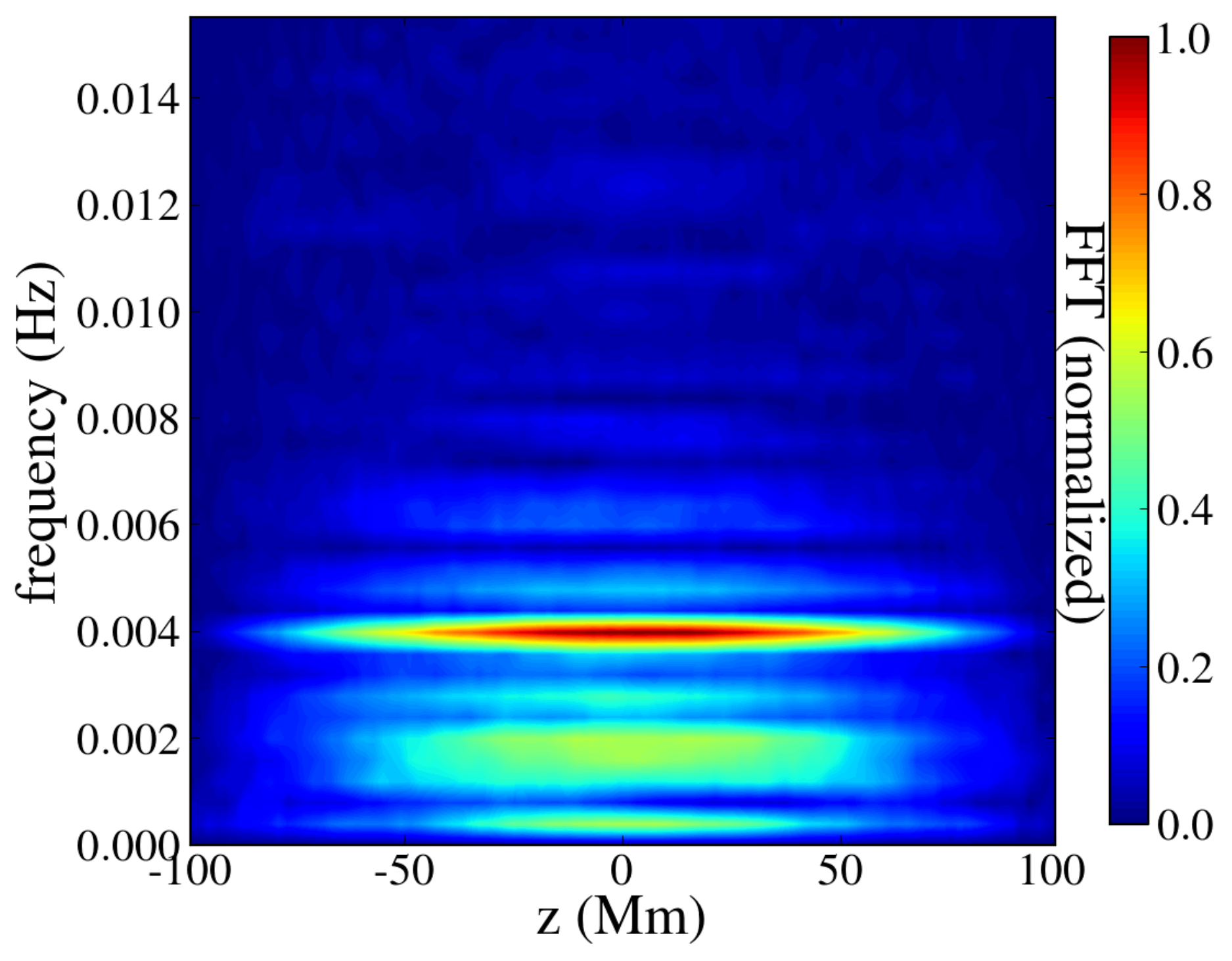}
    \caption{Left panels: displacement of the loop center of mass at each height, for the duration of the simulation. The top panel is for the full time-series, and the bottom panel is for the detrended time series. Right panels: the corresponding normalized power spectral density plots for the two signals.}
    \label{fig:spectra}
\end{figure*}

In \citet{gruszecki2010}, a bluff body was used, which would interact with the flow, but would not be affected by it. For our 3D setup, where our obstacle is not a bluff body but a loop allowed to oscillate, we expected that the background flow will deform the initial circular loop cross-section. Indeed, this can be seen in the panels of Figure \ref{fig:density}, where the first six snapshots of the simulation are shown, between $t=0$ and $t=189.8$\,s, for every $31.62$\,s. Eventually, vortex shedding is initiated, as we can see for the density and $z$-vorticity in Figure \ref{fig:vort}, for snapshots at $t=1012$, $1265$ and $1518$\,s. Although vortex shedding is indicated from the evolution of the velocity field and the vorticity, the loop cross-section only shows vague signs of displacement in the $y$ direction, perpendicular to the flow. 

To test whether vortex shedding can initiate an oscillation, we tracked the center of mass of the loop cross-section at the $xy$-plane at every height along the $z$ axis. The results are plotted in the top left panel of Figure \ref{fig:spectra}, where the temporal evolution of the loop displacement along the $y$-direction is shown along the loop length. We observe a clear oscillatory pattern, which provides us with the first proof-of-concept for the validity of the mechanism proposed in \citet{nakariakov2009}. This oscillatory pattern occurs on top of a mean displacement of the loop center of mass at the later stages of the simulation. By plotting the normalized power spectral density along the loop in the top right panel of Figure \ref{fig:spectra}, we can identify the spatial and temporal harmonic structure of our oscillator. This height-frequency ($z-f$) diagram shows a local maximum $\sim 0.004$\,Hz, which is the eigenfrequency of the fundamental kink mode for our loop $f_0=1/P\sim0.0039$\, where $P=253$\,s. Due to the limits imposed by the relatively short time series, and the overall broadband nature of our driving mechanism, many additional frequencies are shown to be excited. We can see the inclusion of some very low frequencies presumably due to the measured mean displacement of the loop center of mass. Identifying some of these additional frequencies in a future study could be very useful for coronal seismology.

\begin{figure}[t]
    \centering
    \includegraphics[trim={0.0cm 0.cm 0.0cm 0.cm},clip,scale=0.4]{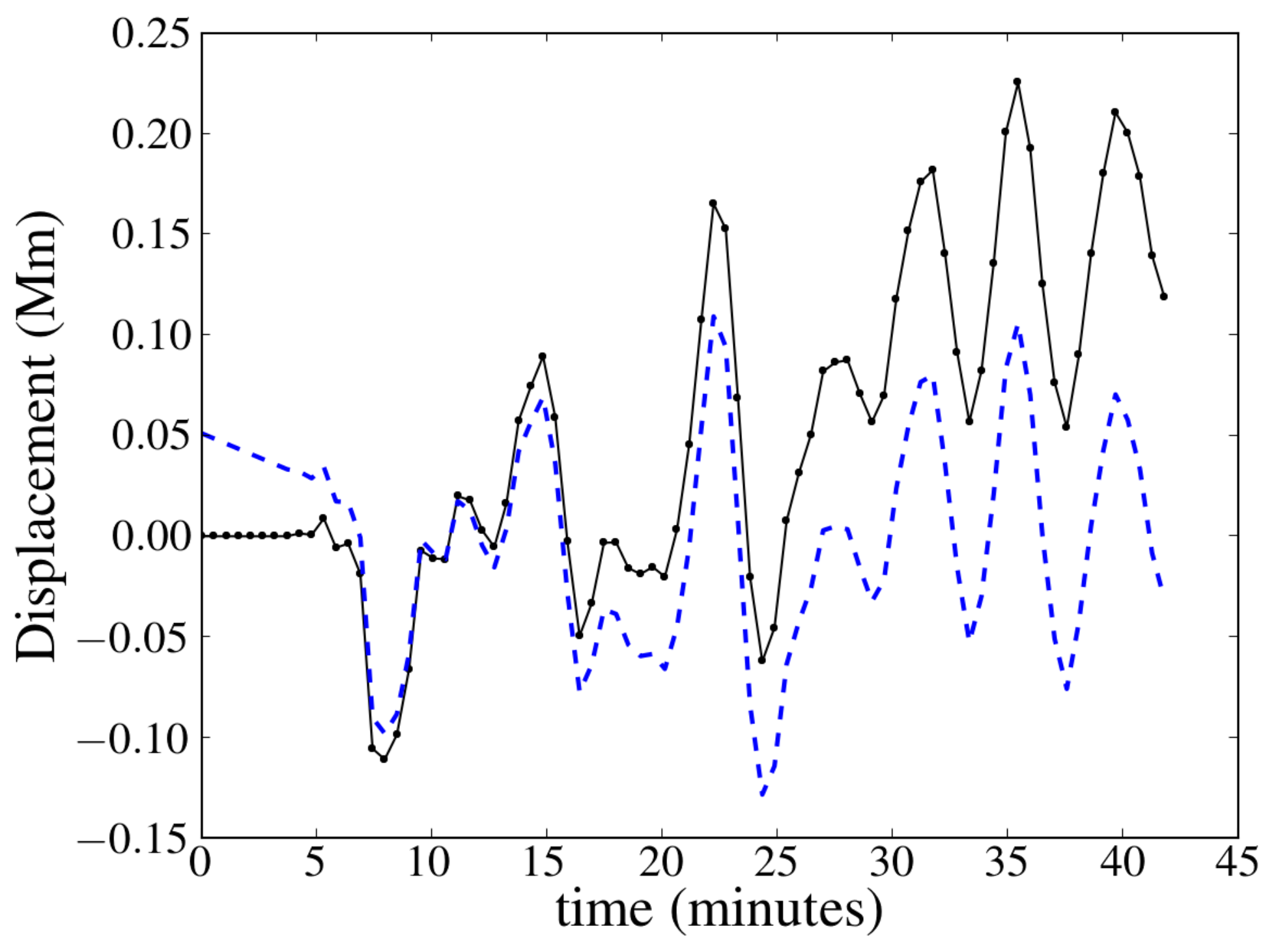}
    \caption{Solid black line: displacement of the center of mass at the apex, in the y-direction. Dashed blue line, the oscillating signal after the mean displacement is subtracted.}
    \label{fig:apex}
\end{figure}

In order to remove the effects of the loop mean displacement along the $y$ direction from our power spectra density plots, we detrend our time-series using the \textit{scipy.signal.detrend} command for Python. This performs a linear least-squares fit to the data, and then subtracts that result (i.e. the mean displacement) from the initial data (i.e. the overall displacement). As we can see from Figure \ref{fig:apex} for the displacement of the center of mass across the background flow at the apex, the signal is detrended once the oscillation starts, although we end up with falsely pronounced values at the very beginning of the simulation.
%
%In order to remove the effects of the loop mean displacement along the $y$ direction from our power spectra density plots, we tried to subtract this mean displacement from the overall loop displacement. To do that, we use a simple approach, by using the \textit{scipy.signal.detrend} command from the SciPy package for Python. \textbf{The default function of this command is to perform a linear least-squares fit to the initial data, and then subtract the result (i.e. the mean displacement) from the initial data (i.e. the overall loop displacement).} As we can see from Figure \ref{fig:apex} for the displacement of the centre of mass across the background flow at the apex, this command manages to detrend the signal once the oscillation starts, although it gives us falsely pronounced values at the very beginning of the simulation. 

Applying that detrending to the entire time series we get the detrended displacement of the entire loop over time, at the bottom left panel of Figure \ref{fig:spectra}. As we see from that panel, the oscillation amplitudes on the detrended signal are of the order of $0.1$\,Mm, which are comparable with those of the observed decay-less oscillations \citep{anfinogentov2015}. This brings the 0D model of \citet{nakariakov2009} to the level of 3D simulations. In addition, we do not see an obvious and consistent decay of this amplitude over time, due to the continuous presence of the background flow. Despite the simplicity of this model, this agreement indicates that vortex shedding can potentially be a mechanism sustaining decay-less oscillations in loops. From the power spectra on the bottom right panel of the same figure, we can clearly see that the frequency of the fundamental kink mode is the one excited the most. Finally, it is safe to assume that once the background flow weakens substantially, vortex shedding will not be able to sustain the oscillation and the amplitude will decay.

\begin{figure}[t]
    \centering
    \includegraphics[trim={0.cm 0.cm 0.cm 0.cm},clip,scale=0.45]{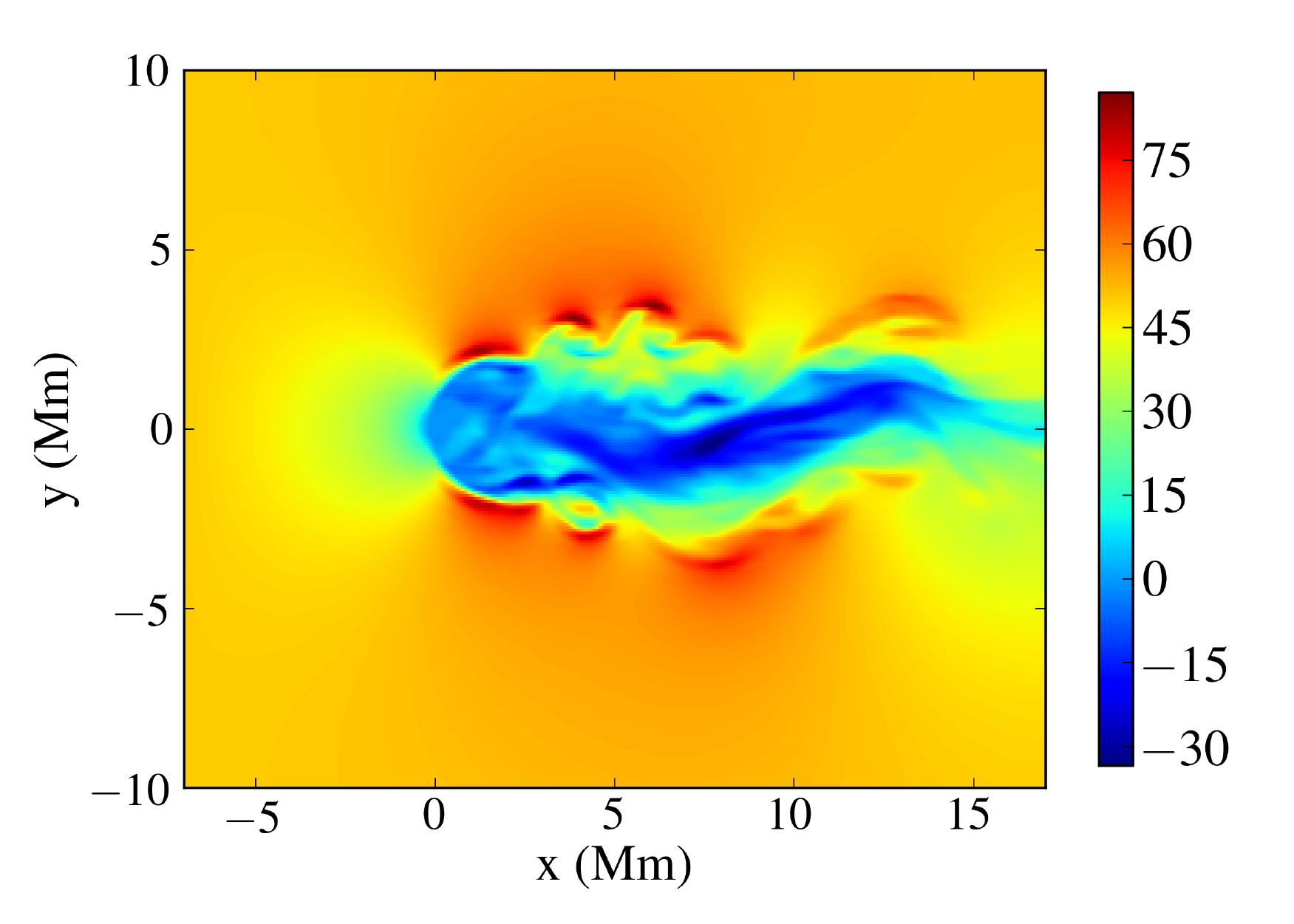}
    \caption{Profile of the $v_x$ ($\times 10^3$\,m s$^{-1}$) velocity at the apex, for part off our datacube depicting the loop at $t=1012$\,s.}
    \label{fig:velocity}
\end{figure}

In the same figure, we see that there is an additional band of frequencies near $f=2$\,mHz in the normalized power spectral density plot. From the equation of the Strouhal number, and assuming that its values are $St\sim 0.2$, this frequency can be obtained for velocities of the order of $20 \times 10^4$\,m\,s$^{-1}$. As we can see from the velocity profile at the apex (Figure \ref{fig:velocity}; also hinted from the velocity field in Figure \ref{fig:vort}), the velocity field in front of the loop drops to values around $20-25$\,km\,s$^{-1}$. This could potentially explain the peak near $2$\,mHz, as the frequency imposed by the flow. Additionally, the fact that the loop cross-section changes over time will inevitably affect the frequency imposed by vortex shedding, explaining the width of the frequency band around $2$\,mHz. Also, it is possible that the value of the Strouhal number in 3D environments in the presence of magnetic field needs is different from the one calculated in \citet{gruszecki2010} for the 2D case. A full 3D parameter study was outside the scope of this work, and therefore not addressed here. However, the aforementioned frequency band near $2$\,mHz seems to be dictated by the Strouhal number of the flow, as shown in the following paragraph.

\begin{figure}[t]
    \centering
    \includegraphics[trim={0.cm 0.cm 0.cm 0.cm},clip,scale=0.4]{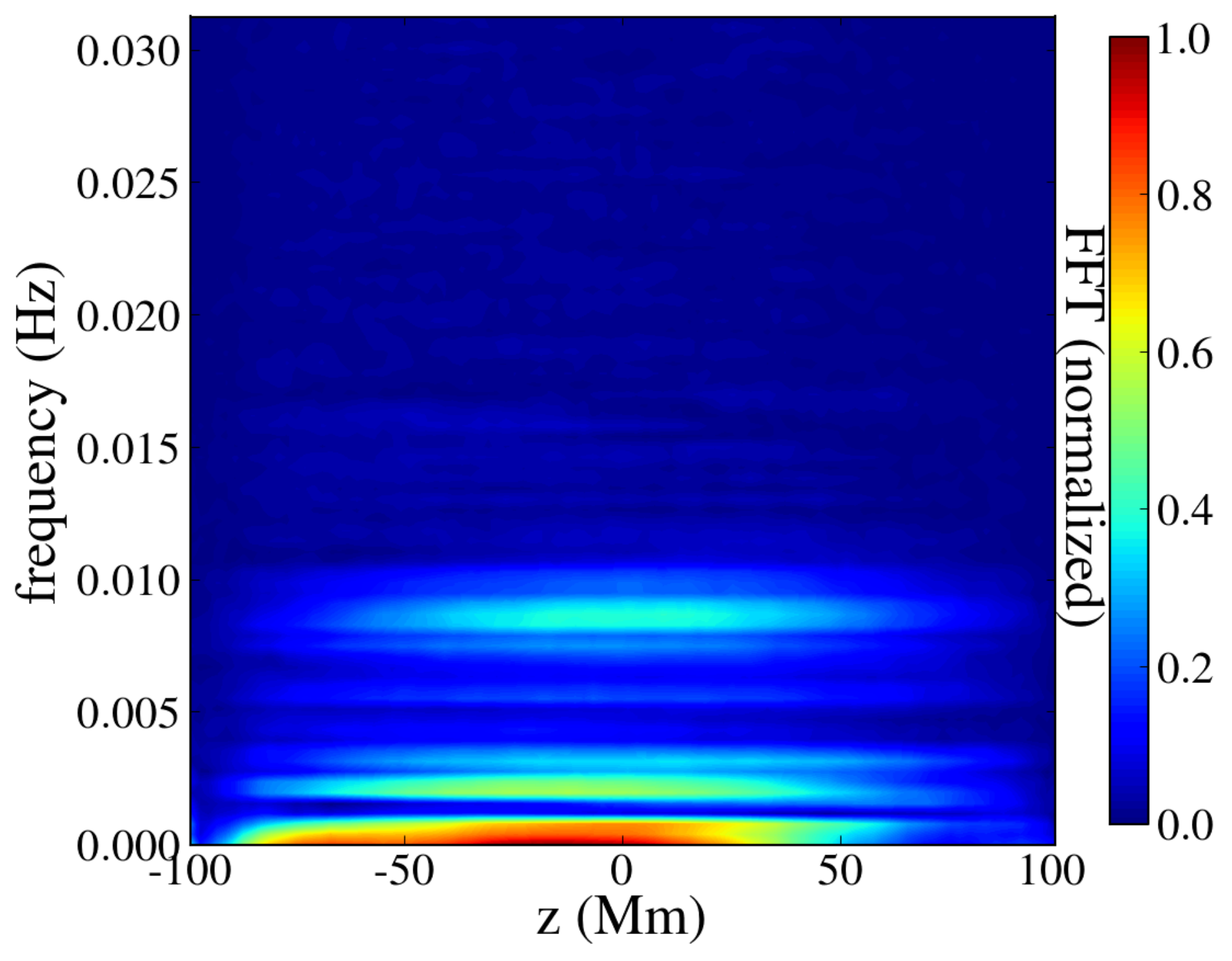}
    \caption{Plot of the normalized power spectral density for the oscillation of the loop with internal and external magnetic field values of $B_{zi} =45.6$\,G and $B_{ze} = 45.64$\,G (secondary setup). }
    \label{fig:spectra2}
\end{figure}

As the final step, we want to test whether driving an oscillation via vortex shedding is a self-oscillating process. For that, we have considered a secondary setup, identical to the one described in section \ref{sec:setup}, but with internal and external magnetic field values of $B_{zi} =45.6$\,G and $B_{ze} = 45.64$\,G, respectively, doubling the Alfv\'{e}n speed and halving the period of the fundamental kink oscillation. In this new setup, we chose to drive the background flow for the whole duration of the simulation $\Delta t = 2536$\,s, because the stronger magnetic field makes the loop ``stiffer,'' delaying the start of the oscillation. This new loop has a fundamental kink mode frequency of $\sim7.9$\,mHz, represented by a strong peak in the normalized power spectral density plot for its nondetrended oscillating  displacement, seen in Figure \ref{fig:spectra2}. This peak is different from the one dictated by the Strouhal number of the flow, which has the same velocity as before. The strong signal at near-zero frequencies due to the loop mean displacement is also shown here, as was in the case in the nondetrended signal from Figure \ref{fig:spectra}. A wide frequency band between $2$ and $4$\,mHz is also visible in this setup, showing that it is indeed dictated by the Strouhal number of the flow, as was previously mentioned, rather than the loop. The fact that the loop eigenfrequency is prominent, while the flow velocity is the same as before, shows that the loop is amplifying or imposing its own preferred frequency. Additionally, since the driving comes from an initially uniform and steady background flow, this process fits the definition of a self-oscillation \citep{jenkins2013PhR}. 

\section{Summary and Conclusion}
Although the idea that vortex shedding can excite standing waves in coronal loops has been proposed in \citet{nakariakov2009}, this mechanism had not been properly explored in a full 3D MHD model. The phenomenon of Alfv\'{e}nic vortex shedding has already been studied for a bluff body in 2D by \citet{gruszecki2010}. However, a coronal loop will interact differently with a background flow than a fixed and rigid bluff body, and thus a study in 3D was essential. In the current work we see that a nonrigid loop will be deformed when interacting with a background flow. In addition, we see for the first time in a 3D simulation that the vortex shedding will eventually force the loop into an oscillation in the direction perpendicular to the flow, as was first proposed in \citet{nakariakov2009}. 

The long duration of the vortex-shedding driving and the oscillation amplitudes comparable to the those found in \citet{anfinogentov2015} make this mechanism a good candidate for generating decay-less oscillations. However, once the background flow is no longer present, the oscillations would start decaying. Thus, also decaying oscillations could potentially be generated via vortex shedding for short-lived background flows.

The mechanism of vortex shedding induced oscillations seems to fall into the self-oscillation processes. These, as was described in \citet{jenkins2013PhR}, are processes that can turn a nonperiodic driving (like a background steady flow) into a periodic signal (like a loop oscillation). Although vortex shedding has a preferred periodicity, as described by the flow Strouhal number \citep{gruszecki2010}, the spectral densities of two different loops have clearly shown strong peaks around the corresponding frequencies of their fundamental standing kink-modes and not near the peak dictated by the Strouhal number. This shows that the oscillating loop imposes its own frequency, which is a characteristic of self-oscillations. Future studies could also help identify some of the additional frequencies and harmonics observed, which could be important for coronal seismology.

Despite its successes though, our very simple model still only provides a proof-of-concept for sustaining decay-less oscillations through the vortex shedding, and additional studies are necessary. Although the results of \citet{gruszecki2010} for the Strouhal number seem to match the hydrodynamical case, a full parameter study in 3D MHD should be performed, as this was outside the scope of the current study. In order to compare with observations, gravitationally stratified loops and realistic flow profiles should be considered. Changing the flow speed would change the location of frequency peak dictated by the Strouhal number, but it could also have catastrophic effects on the loop cross-section, should the flow be too strong or the loop not ``stiff'' enough (i.e. having a weaker magnetic field). In addition, the loop characteristics are expected to affect the resulting oscillation amplitudes, and thus the strength of the different frequency peaks. %However, in this work we have clearly shown that vortex shedding can be generated from a background flow in the presence of a coronal loop, forcing the loop into a self-oscillation, and giving a candidate mechanism for explaining undamped transverse waves in coronal loops. 

\acknowledgments
The authors would like to thank the referee for their helpful comments. K.K. has received support for this study through a postdoctoral mandate from KU Leuven Internal Funds (PDM/2019), by a UK Science and Technology Facilities Council (STFC) grant ST/T000384/1, and by a FWO (Fonds voor Wetenschappelijk Onderzoek – Vlaanderen) postdoctoral fellowship (1273221N). T.V.D. is supported by the European Research Council (ERC) under the European Union's Horizon 2020 research and innovation program (grant agreement No. 724326) and the C1 grant TRACESpace of Internal Funds KU Leuven. The computational resources and services used in this work were provided by the VSC (Flemish Supercomputer Center), funded by the Research Foundation Flanders (FWO) and the Flemish Government – department EWI.

\bibliography{paper}{}
\bibliographystyle{aasjournal}

\end{document}